\documentclass[12pt]{article}
\pdfoutput=1
\usepackage{hyperref}
\usepackage{amsmath,amssymb,amsthm,amscd}
\usepackage{mathtools}
\usepackage{bm}
\usepackage{graphicx}

\newcommand{\Z}{\mathbb{Z}}

\newcommand{\Tr}{\textrm{Tr}}

\newcommand{\be}{\begin{equation}}
\newcommand{\ee}{\end{equation}}
\newcommand{\ba}{\begin{aligned}}
\newcommand{\ea}{\end{aligned}}

\numberwithin{equation}{section}

\begin{document}
\hypersetup{pageanchor=false}
\begin{titlepage}

\rightline{USTC-ICTS/PCFT-26-69}

\vskip 3 cm

\begin{center}
{\Large\bfseries Algebraic Proofs of\\[0.15em]
Quantum Period--TBA Relations\par}
\end{center}

\vskip 0.5 cm

\renewcommand{\thefootnote}{\fnsymbol{footnote}}
\vskip 30pt \centerline{ {\large \rm 
Min-xin Huang\footnote{minxin@ustc.edu.cn}  
} } \vskip .5cm  \vskip 20pt 

\begin{center}
{Interdisciplinary Center for Theoretical Study,  \\ \vskip 0.1cm  University of Science and Technology of China,  Hefei, Anhui 230026, China} 
 \\ \vskip 0.3 cm
{Peng Huanwu Center for Fundamental Theory,  \\ \vskip 0.1cm  Hefei, Anhui 230026, China} 
\end{center}

\setcounter{footnote}{0}
\renewcommand{\thefootnote}{\arabic{footnote}}
\vskip 40pt
\begin{abstract}

We give algebraic proofs of two conjectural relations between quantum
$A$-periods and thermodynamic Bethe ansatz (TBA)-like equations for quantum
mirror curves. The quadratic relation follows from constant terms of
shifted product expansions. For higher-degree relations, a cyclic
quantum-torus trace and a finite matrix determinant identify the period
with the free energy of nonoverlapping rods. Their density equation yields
the TBA-like equation, with normalization fixed by the rod length.
Both proofs hold as formal power series in the complex structure
parameter, with exact dependence on $q=e^{i\hbar}$.

\end{abstract}

\end{titlepage}
\hypersetup{pageanchor=true}
\vfill \eject


\newpage

\baselineskip=16pt

\tableofcontents

\section{Introduction}\label{sec:introduction}

With the assistance of AI tools,
we study the relations between quantum periods and thermodynamic Bethe
ansatz (TBA)-like equations proposed in \cite{Du:2020nwl}, and give
algebraic proofs of two identities that were previously tested in
perturbative expansions. These identities concern quantum $A$-periods,
or quantum mirror maps, and can be formulated without choosing a
Hilbert-space realization of the quantum curve.

Period integrals play a central role in mirror symmetry. The classical
mirror map relates complex structure parameters of a Calabi--Yau
manifold to the K\"ahler parameters of its mirror, allowing period
calculations to determine enumerative information, as illustrated by
the celebrated work of Candelas, de la Ossa, Green and Parkes
\cite{Candelas:1990rm}. For local toric Calabi--Yau threefolds, much of
this information is encoded in a mirror curve. Promoting its coordinates
to operators satisfying $[\hat x,\hat p]=i\hbar$ leads to a difference
equation and a quantum deformation of its periods
\cite{Aganagic:2011mi}. The parameter $\hbar$ is related to the surviving
deformation parameter in the Nekrasov--Shatashvili limit of the
$\Omega$ background, which connects supersymmetric gauge theories to
quantum integrable systems \cite{Nekrasov:2009rc}. Quantum periods also
provide an effective way to compute topological string amplitudes in
this limit \cite{Huang:2012kn}.

These developments have important consequences for spectral theory.
Quantum periods enter the semiclassical quantization of operators
obtained from mirror curves, while their exact spectra require
additional nonperturbative information \cite{Huang:2014eha}.
The spectral theory/topological string correspondence proposed by
Grassi, Hatsuda and Mari\~no \cite{Grassi:2014zfa}, together with the
exact quantization condition \cite{Wang:2015wdy}, gives precise relations between this information
and topological string free energies. Extensions to cluster integrable
systems and higher-genus mirror curves exhibit the role of several
commuting Hamiltonians and independent periods
\cite{Franco:2015rnr,Huang:2020neq}. Moriyama's study of del Pezzo
geometries further shows how exceptional Weyl symmetries organize
quantum curves and quantum mirror maps \cite{Moriyama:2020lyk}.

The TBA description provides another way to access the same class of
quantum problems. The relation between integral-operator resolvents
and nonlinear TBA equations has its roots in the work of
Zamolodchikov \cite{Zamolodchikov:1994uw} and the proofs of Tracy and
Widom \cite{Tracy:1995ax}. In the Fermi-gas formulation of ABJM theory
developed by Mari\~no and Putrov \cite{Marino:2011eh}, this structure
became a powerful tool for exact partition functions and instanton
calculations \cite{Hatsuda:2012hm,Putrov:2012zi,Calvo:2012du}.
A particularly insightful step was the identification by Hatsuda,
Mari\~no, Moriyama and Okuyama \cite{Hatsuda:2013oxa} of the connection
between the ABJM membrane-instanton expansion, refined topological
strings, and quantum periods of local $\mathbb P^1\times\mathbb P^1$.
Their TBA formulation exposed a striking relation between a nonlinear
functional equation and the quantum mirror map. K\"all\'en and
Mari\~no \cite{Kallen:2013qla} gave a compelling spectral explanation
of this connection and demonstrated the essential role of corrections
nonperturbative in $\hbar$ in the quantization problem.

An elegant generalization was developed by Okuyama and Zakany
\cite{Okuyama:2015pzt}, who constructed TBA-like systems for quantized
mirror curves, with local $\mathbb P^2$ as a main example. Their use of
resolvent decompositions and quantum Wronskian relations supplied a
systematic route beyond the original ABJM kernel. The integral kernels
of Kashaev and Mari\~no \cite{Kashaev:2015kha}, expressed through
Faddeev's quantum dilogarithm \cite{Faddeev:1993rs}, were a key
ingredient. These constructions revealed a rich structure that is far
from apparent in a direct expansion of the wavefunction. The subsequent
mass-dependent extension for local $\mathbb P^1\times\mathbb P^1$
\cite{Huang:2020yli} and the analysis of three-term operators in
\cite{Du:2020nwl} led to the two families of period relations considered
here.

There is also a complementary connection between TBA equations and
exact WKB analysis. Ito, Mari\~no and Shu \cite{Ito:2018eon} derived
TBA equations for resummed WKB periods of polynomial Schr\"odinger
problems, with the Stokes discontinuities supplying the relevant
analytic data. Related developments include the deformation of quantum
mechanics studied by Grassi and Mari\~no \cite{Grassi:2018bci}, the
analysis of deformed supersymmetric quantum mechanics by Ito and Shu
\cite{Ito:2024nlt}, and the TBA description of quantum periods in
$D$-type Argyres--Douglas theories by Ito and Yang \cite{Ito:2024wxw}.
The work of Shu, Ito and Yang on wall crossing
\cite{Shu:2025mwe} further illustrates how analytic continuation
changes the TBA system. In the present paper, the term TBA-like refers
specifically to functional difference equations solved as formal
series in a complex structure parameter $z$, with $q=e^{i\hbar}$
kept exact. Identifying these formal solutions with physical integral
equations requires additional analytic information.

Quantum periods have recently acquired a direct enumerative
interpretation. Gr\"afnitz, Ruddat, Zaslow and Zhou
\cite{Grafnitz:2025anq} identify, for the Fano surfaces in their
construction, a refined open mirror map defined by quantum periods
with a quantum theta function whose coefficients encode all-genus
two-point logarithmic Gromov--Witten invariants. They also establish
the equivalence between constant-term and wavefunction definitions
of the quantum $A$-period. Their results include local $\mathbb P^2$,
the $m=1$ case of our three-term family. The relation to ordinary open
Gromov--Witten invariants involves a nontrivial higher-genus correction,
so these different enumerative interpretations must be distinguished.

Our purpose is to establish the two formal period identities directly
from their defining equations. With
$\operatorname{CT}_X f(X)=[X^0]f(X)$ and $\theta_z=z\partial_z$, they are
\[
 \operatorname{CT}_X\eta_s(X)=1+2\theta_z\Pi_s(z),
 \qquad
 \operatorname{CT}_X\eta_m(X)=1+(m+2)\theta_z\Pi_m(z).
\]
Here $\eta_s$ and $\eta_m$ solve the quadratic and higher-degree
TBA-like equations, respectively, and $\Pi_s,\Pi_m$ denote the
quantum-period corrections with the classical logarithmic term
removed. The first identity holds for the general Laurent-series
coefficient $s(X)$ specified in Section~\ref{sec:quadratic}; the second
holds for every positive integer $m$ in the family
\eqref{eq:second-curve}. For the corresponding specialization of $s$,
the two periods coincide, even though the two TBA functions differ.

The two proofs use different algebraic features of the problem.
For the quadratic relation, a $q$-difference Riccati equation produces
forward- and backward-shifted product expansions. Their constant terms
agree by a change of the Laurent variable. For the higher-degree
relation, a cyclic constant-term trace on the quantum torus expresses
the period as a trace of a logarithm. A change of quantum variables then
turns a finite band determinant into a partition function for
nonoverlapping rods of length $m+2$. The density equation of these rods
is precisely the required TBA-like equation. This connects the proof
to the exclusion-statistics formalism of Ouvry and Polychronakos
\cite{OuvryPolychronakos2019,OuvryPolychronakos2021} and to the
lattice-walk approach to quantum periods \cite{Gan2025}. In particular,
the factor $m+2$ follows from the rod length. All limits and identities
are taken coefficientwise in $z$; no convergence assumption or
expansion in $\hbar$ enters the proofs.

The paper is organized as follows. Section~\ref{sec:quadratic} proves
the quadratic relation and explains its Riccati form.
Section~\ref{sec:higher} develops the quantum-torus and lattice
construction and proves the higher-degree relations.
Section~\ref{sec:conclusion} discusses the physical and geometric
questions suggested by these results.

\section{The quadratic TBA relation}\label{sec:quadratic}

The quadratic TBA relation, equation~(3.24) of \cite{Du:2020nwl}, is an identity between the constant terms of a quantum-period integrand and a solution of a TBA-like equation.  We write
\begin{equation}
 \operatorname{CT}_X f(X)=[X^0]f(X)
 =\operatorname{Res}_{X=0}\frac{f(X)}{X}.
 \label{eq:first-ct}
\end{equation}
All the statements below are identities of formal power series in $z$, with $q\ne0$ kept exact.  More precisely, their coefficients are Laurent series in $X$ with a finite lower bound on the powers of $X$ at each order in $z$.  This makes multiplication and constant-term extraction well defined.  No convergence assumption on $z$ or expansion in $\hbar$, where $q=e^{i\hbar}$, is needed.

To recall the normalization, let $m$ be an integer and $r(X)$ a nonzero rational function. The quantum curve
\begin{equation}
 \bigl(e^{\hat p}+z e^{-m\hat x-\hat p}-r(e^{\hat x})\bigr)\psi(x)=0,
 \qquad [\hat x,\hat p]=i\hbar,
 \label{eq:first-curve}
\end{equation}
with $X=e^x$ gives the difference equation
\begin{equation}
 \frac{1}{V(X)}+zq^{-m/2}X^{-m}V(qX)=r(X),
 \qquad V(X)=\frac{\psi(x)}{\psi(x-i\hbar)}.
 \label{eq:first-original-v}
\end{equation}
Here and below the parameters $z,q$ in $V(X;q,z)$ are usually left implicit.  The branch relevant to the quantum period has $V(X)=r(X)^{-1}+O(z)$.  The symmetrically shifted normalization
\begin{equation}
 \widetilde V(X)=r(q^{-1/2}X)V(q^{-1/2}X),
 \qquad
 s(X)=\frac{1}{X^m r(q^{1/2}X)r(q^{-1/2}X)}
 \label{eq:first-normalization}
\end{equation}
then obeys
\begin{equation}
 \frac{1}{\widetilde V(X)}+zs(X)\widetilde V(qX)=1,
 \qquad \widetilde V(X)=1+O(z).
 \label{eq:first-riccati}
\end{equation}
In fact, the proof applies to any Laurent series $s(X)$ with a finite lower bound on its powers of $X$, whether or not it is written in the form~\eqref{eq:first-normalization}. We assume that $s$ is independent of $z$, but allow arbitrary dependence on $q$. The parameter $q$ is kept fixed and is also suppressed in the notation for the periods.

Equation~\eqref{eq:first-riccati} has the form of a $q$-difference Riccati
equation: it relates shifted values by a fractional-linear transformation,
\[
 \widetilde V(X)=\frac{1}{1-zs(X)\widetilde V(qX)}.
\]
Introducing an auxiliary sequence $u_n$ by
$\widetilde V(q^nX)=u_n/u_{n-1}$ converts it into the second-order linear
difference equation
\[
 zs(q^nX)u_{n+1}-u_n+u_{n-1}=0.
\]
This is the discrete counterpart of the logarithmic-derivative substitution
that relates a Riccati differential equation to a second-order linear
differential equation; see Section~2 of \cite{Huang2013Riccati}.
A difference-Riccati formulation of quantum mirror curves is also used
in the WKB analysis of Del Monte and Longhi
\cite[Section~3.5]{DelMonteLonghi2023}.

Let $\eta_s(X)=1+O(z)$ be the solution of
\begin{equation}
 \eta_s(X)^2=1+zs(X)
 \bigl[\eta_s(qX)+\eta_s(X)\bigr]
 \bigl[\eta_s(q^{-1}X)+\eta_s(X)\bigr],
 \label{eq:first-tba}
\end{equation}
and define the quantum correction to the period by
\begin{equation}
 \Pi_s(z)=\operatorname{CT}_X\log\widetilde V(X),
 \qquad \theta_z=z\frac{\partial}{\partial z}.
 \label{eq:first-period}
\end{equation}
The claim is
\begin{equation}
 \boxed{\operatorname{CT}_X\eta_s(X)
       =1+2\theta_z\Pi_s(z).}
 \label{eq:first-relation}
\end{equation}
Both equations~\eqref{eq:first-riccati} and~\eqref{eq:first-tba} determine their indicated branches uniquely, order by order in $z$.  For the former, use $\widetilde V(X)=1+zs(X)\widetilde V(X)\widetilde V(qX)$; for the latter, the coefficient of $z^n$ in $\eta_s(X)^2$ contains twice the unknown coefficient of $\eta_s(X)$, while the remaining terms involve lower orders.

The proof uses two expansions which differ only in the direction of their $q$ shifts.  Set
\begin{equation}
 R(X)=\widetilde V(X)-1.
 \label{eq:first-r}
\end{equation}
Equation~\eqref{eq:first-riccati} becomes
\begin{equation}
 R(X)=zs(X)\bigl[1+R(X)\bigr]\bigl[1+R(qX)\bigr].
 \label{eq:first-r-equation}
\end{equation}
Consider the backward-shifted product sum
\begin{equation}
 F(X)=\sum_{k\ge1}\prod_{j=0}^{k-1}R(q^{-j}X).
 \label{eq:first-f}
\end{equation}
Since $R(X)=O(z)$, only $k\le n$ can contribute to the coefficient of $z^n$.  Thus this sum is a well-defined formal series.  Separating its first factor gives
\begin{equation}
 F(X)=R(X)\bigl[1+F(q^{-1}X)\bigr].
 \label{eq:first-f-recursion}
\end{equation}
Consequently,
\begin{align}
 1+F(X)+F(q^{-1}X)
  &=\bigl[1+R(X)\bigr]\bigl[1+F(q^{-1}X)\bigr],\notag\\
 1+F(X)+F(qX)
  &=\bigl[1+R(qX)\bigr]\bigl[1+F(X)\bigr].
 \label{eq:first-f-sums}
\end{align}
Multiplying these identities and using~\eqref{eq:first-r-equation} and~\eqref{eq:first-f-recursion}, we obtain
\begin{equation}
 zs(X)\bigl[1+F(X)+F(q^{-1}X)\bigr]
        \bigl[1+F(X)+F(qX)\bigr]
 =F(X)\bigl[1+F(X)\bigr].
 \label{eq:first-f-tba}
\end{equation}
It follows that $1+2F(X)$ satisfies~\eqref{eq:first-tba} and has the correct leading term.  Uniqueness therefore identifies the TBA solution explicitly as
\begin{equation}
 \eta_s(X)=1+2F(X)
 =1+2\sum_{k\ge1}\prod_{j=0}^{k-1}R(q^{-j}X).
 \label{eq:first-eta-products}
\end{equation}

To obtain the other product sum, differentiate~\eqref{eq:first-riccati} with respect to $\log z$.  Equivalently, differentiate
$\widetilde V(X)=1+zs(X)\widetilde V(X)\widetilde V(qX)$ and divide by $\widetilde V(X)$.  Using $zs(X)\widetilde V(qX)=1-\widetilde V(X)^{-1}$ gives
\begin{equation}
 \theta_z\log\widetilde V(X)
 =R(X)\bigl[1+\theta_z\log\widetilde V(qX)\bigr].
 \label{eq:first-derivative-recursion}
\end{equation}
Iteration now proceeds in the opposite direction:
\begin{equation}
 \theta_z\log\widetilde V(X)
 =\sum_{k\ge1}\prod_{j=0}^{k-1}R(q^jX).
 \label{eq:first-forward-products}
\end{equation}
Finally, constant-term extraction is invariant under $X\mapsto q^\ell X$ for every integer $\ell$.  At each fixed $k$, this gives
\begin{align}
 \operatorname{CT}_X\prod_{j=0}^{k-1}R(q^jX)
 &=\operatorname{CT}_X\prod_{j=0}^{k-1}R(q^{j-k+1}X)\notag\\
 &=\operatorname{CT}_X\prod_{j=0}^{k-1}R(q^{-j}X).
 \label{eq:first-reverse-products}
\end{align}
The last equality simply reverses the order of scalar factors.  Taking constant terms in~\eqref{eq:first-eta-products} and~\eqref{eq:first-forward-products} proves~\eqref{eq:first-relation} to every order in $z$.

As a classical check, when $s$ has a well-defined specialization at
$q=1$, the two defining equations give
\[
 \widetilde V(X)=\frac{2}{1+\sqrt{1-4zs(X)}},
 \qquad \eta_s(X)=\frac{1}{\sqrt{1-4zs(X)}}.
\]
Here all functions are evaluated at $q=1$, and the square root has
constant term one in $z$. Direct differentiation gives
$\eta_s(X)=1+2\theta_z\log\widetilde V(X)$ pointwise, checking
\eqref{eq:first-relation} before constant-term extraction.

Both defining equations also occur in the exclusion-statistics
formalism of Ouvry and Polychronakos \cite{OuvryPolychronakos2021}.
For exclusion parameter two, their forward effective single-level
partition function obeys $y_i=1+w_i/y_{i+1}$, equation~(21) of the
published paper. Setting
\[
 w_i=-zs(q^iX),\qquad y_i=\widetilde V(q^iX)^{-1}
\]
gives precisely the Riccati equation~\eqref{eq:first-riccati}.
Their density equation~(25) becomes~\eqref{eq:first-tba} with
occupation $n_i=(1-\eta_s(q^iX))/2$.
These statistical identities underlie the correspondence; the proof
above identifies the associated quantum-period relation by
constant-term extraction. The next section uses the corresponding
determinant and density constructions
\cite{OuvryPolychronakos2019,OuvryPolychronakos2021}
with exclusion parameter $m+2$.

For the curve studied in the next section, $r(X)=1-X$ and
\begin{equation}
 s(X)=\frac{X^{-m}}{(1-q^{1/2}X)(1-q^{-1/2}X)}.
 \label{eq:first-special-s}
\end{equation}
Since $\operatorname{CT}_X\log(1-X)=0$ and constant terms are invariant under the half shift as well, the above period equals
\begin{equation}
 \Pi_s(z)=\operatorname{CT}_X\log\bigl[(1-X)V(X)\bigr]
          =\operatorname{CT}_X\log V(X)
          =\Pi_m(z).
 \label{eq:first-special-period}
\end{equation}
Here $\log V(X)$ is understood as its $z=0$ value $-\log(1-X)$ plus the formal $z$-series $\log[(1-X)V(X)]$.  Thus the two TBA relations considered here refer to the same quantum-period correction, although their TBA-like equations are different.

\section{Higher-degree TBA relations}\label{sec:higher}

Throughout this section, $m$ is a positive integer. We prove equation (4.28)
of \cite{Du:2020nwl} as an identity of formal power series in $z$, with the
dependence on $q$ kept exact. All residues below mean constant terms in the
Laurent expansion at $X=0$, with the same convention as in the preceding
section.

\subsection{The equations and the quantum period}

The quantum mirror curve and its equation for the wave-function ratio are
\begin{equation}\label{eq:second-curve}
 \bigl(e^{\hat x}+e^{\hat p}+z e^{-m\hat x-\hat p}\bigr)\psi(x)=\psi(x),
 \qquad [\hat x,\hat p]=i\hbar,
\end{equation}
\begin{equation}\label{eq:second-V}
 \frac{1}{V(X)}+zq^{-m/2}X^{-m}V(qX)=1-X,
 \qquad V(X)=\frac{1}{1-X}+O(z).
\end{equation}
Here $X=e^x$ and $q=e^{i\hbar}$. As in \cite{Du:2020nwl}, the notation
$V(X)$ suppresses the parameters $q,z$; we write $V(X;q,z)$ when a change
of these parameters must be displayed. The period is
\begin{equation}\label{eq:second-period}
 \Pi_m(z)=\operatorname{CT}_X\log[(1-X)V(X)]
         =\operatorname{CT}_X\log V(X).
\end{equation}
The second equality uses $\operatorname{CT}_X\log(1-X)=0$.

Define the Laurent polynomial
\begin{equation}\label{eq:P-definition}
 P(X)=X^{-m}\prod_{j=0}^{m}(1+q^{j-m/2}X).
\end{equation}
There are $m+1$ factors in this product. For example,
$P(X)=X^{-1}+q^{-1/2}+q^{1/2}+X$ for $m=1$. The second TBA-like equation,
equation (4.26) of \cite{Du:2020nwl}, is
\begin{equation}\label{eq:second-TBA}
 (1-\eta_m(X))\prod_{i=-m}^{0}B(q^iX)
 =-zP(X)\prod_{i=-(m+1)}^{0}A(q^iX),
\end{equation}
where
\begin{equation}\label{eq:second-AB}
 A(X)=\sum_{j=0}^{m+1}\eta_m(q^jX),
 \qquad B(X)=1+\sum_{j=0}^{m}\eta_m(q^jX).
\end{equation}
We choose the solution $\eta_m(X)=1+O(z)$. It is unique: at each order
$z^r$, the new coefficient of $\eta_m(X)$ appears on the left of
\eqref{eq:second-TBA} multiplied by the nonzero constant
$-(m+2)^{m+1}$, while all other terms involve coefficients of lower order.
The function $\eta_m$ in this section is the solution of
\eqref{eq:second-TBA}; it is different from the first TBA function
$\eta_s$ in the preceding section. The identity to be proved is
\begin{equation}\label{eq:second-target}
 \boxed{\operatorname{CT}_X\eta_m(X)=1+(m+2)\theta_z\Pi_m(z).}
\end{equation}

The logarithmic derivative in~\eqref{eq:second-target} suggests a
partition-function interpretation: differentiating its logarithm counts
occupied rods. We shall identify $\Pi_m$ with minus the constant term
of the logarithm of a lattice partition function per site. A change of
quantum variables turns the operator into two allowed lattice moves,
whose determinant describes nonoverlapping rods of length $m+2$.
The TBA function will be one minus $m+2$ times the rod-start density; its
shifted sums then describe empty intervals of possible rod starts.
This connects the two sides of~\eqref{eq:second-target} to the same
partition function.

\subsection{A constant-term trace for the quantum curve}

Introduce invertible quantum variables
\begin{equation}\label{eq:quantum-torus}
 \hat a\hat b=q\hat b\hat a,
 \qquad
 \hat H=\hat a+\hat b+q^{-m/2}\hat a^{-m}\hat b^{-1}.
\end{equation}
One may think of $\hat a=e^{\hat x}$ and $\hat b=e^{\hat p}$.
The Baker--Campbell--Hausdorff formula then gives
$e^{-m\hat x-\hat p}=q^{-m/2}\hat a^{-m}\hat b^{-1}$.
For the proof, only the algebraic commutation relation is needed;
no Hilbert-space representation is assumed. Hats distinguish these
noncommuting variables from the scalar argument $X$.

Write each Laurent polynomial in the order $\hat a^r\hat b^s$ and define
\begin{equation}\label{eq:tau-definition}
 \tau\!\left(\sum_{r,s}f_{r,s}\hat a^r\hat b^s\right)=f_{0,0}.
\end{equation}
This is a constant-term trace, not a Hilbert-space operator trace. Its
cyclicity follows directly from
\begin{align}
 (\hat a^r\hat b^s)(\hat a^u\hat b^v)
   &=q^{-su}\hat a^{r+u}\hat b^{s+v},\nonumber\\
 (\hat a^u\hat b^v)(\hat a^r\hat b^s)
   &=q^{-vr}\hat a^{r+u}\hat b^{s+v}.
\end{align}
Both constant terms vanish unless $u=-r$ and $v=-s$; in that case both
commutation factors are $q^{rs}$. By linearity,
\begin{equation}\label{eq:tau-cyclic}
 \tau(\hat F\hat G)=\tau(\hat G\hat F).
\end{equation}
We extend $\tau$ coefficientwise to formal power series. In particular,
for $\hat F(t),\hat G(t)=1+O(t)$,
\begin{equation}\label{eq:tau-log-additive}
 \tau\log(\hat F\hat G)=\tau\log\hat F+\tau\log\hat G.
\end{equation}
For completeness, cyclicity in the differentiated logarithm series gives
$\partial_t\tau\log\hat F=\tau(\hat F^{-1}\partial_t\hat F)$.
Applying this formula to $\hat F\hat G$ proves that the derivatives of
the two sides of \eqref{eq:tau-log-additive} agree. Both sides vanish at
$t=0$, proving the identity. Thus no commutativity of $\hat F,\hat G$
is required under the trace.

To relate this trace to the period, introduce a scalar formal variable
$t$ and make the substitutions $X=t\hat a$, $z=t^{m+2}$:
\begin{equation}\label{eq:Vt-definition}
 V_t(\hat a):=V(t\hat a;q,t^{m+2}).
\end{equation}
This is the original function evaluated at new arguments, rather than a
new solution with additional independent variables. Functions of
$\hat a$ commute with one another, although they do not in general
commute with $\hat b$. Equation \eqref{eq:second-V} becomes
\begin{equation}\label{eq:Vt-equation}
 V_t(\hat a)^{-1}
 =1-t\hat a-t^2q^{-m/2}\hat a^{-m}V_t(q\hat a).
\end{equation}
The substitution is well defined formally. Indeed, the coefficient of
$z^r$ in $V(X)$ has a pole of order at most $mr$ at $X=0$, as follows
inductively from \eqref{eq:second-V}. After substitution, this contribution
starts at order $t^{2r}$. At each order in $t$, only finitely many Laurent
monomials in $\hat a$ occur, and $V_t(\hat a)=1+O(t)$.

Using $\hat b^{-1}f(\hat a)=f(q\hat a)\hat b^{-1}$, one now obtains
the factorization
\begin{equation}\label{eq:quantum-factorization}
\begin{aligned}
 1-t\hat H={}&
 \left(1-tq^{-m/2}\hat a^{-m}V_t(q\hat a)\hat b^{-1}\right)
 V_t(\hat a)^{-1}
 \left(1-tV_t(\hat a)\hat b\right).
\end{aligned}
\end{equation}
To check the order of the factors, multiplying them gives
\[
 V_t(\hat a)^{-1}-t\hat b
 -tq^{-m/2}\hat a^{-m}\hat b^{-1}
 +t^2q^{-m/2}\hat a^{-m}V_t(q\hat a),
\]
which equals $1-t\hat H$ by \eqref{eq:Vt-equation}.
Every term in the logarithm of the first factor in
\eqref{eq:quantum-factorization} has strictly negative $\hat b$-degree,
and every term in the logarithm of the last factor has strictly positive
$\hat b$-degree. Their traces therefore vanish. Equations
\eqref{eq:tau-log-additive} and \eqref{eq:second-period} imply
\begin{equation}\label{eq:period-trace-log}
 -\tau\log(1-t\hat H)
 =\tau\log V_t(\hat a)
 =\Pi_m(t^{m+2}).
\end{equation}
In the last equality, a monomial $X^jz^r$ becomes
$t^{j+(m+2)r}\hat a^j$; taking the constant term retains precisely
$j=0$.
Equivalently,
\begin{equation}\label{eq:period-moments}
 \begin{aligned}
 &\tau(\hat H^k)=0\qquad\text{if }(m+2)\nmid k,\\
 &\Pi_m(z)=\sum_{r\geq1}
 \frac{z^r}{(m+2)r}\tau\!\left(\hat H^{(m+2)r}\right).
 \end{aligned}
\end{equation}
Indeed, if a word in the expansion of $\hat H^k$ contains $r$ copies
of $q^{-m/2}\hat a^{-m}\hat b^{-1}$, zero total $\hat a$- and
$\hat b$-degrees require respectively $mr$ copies of $\hat a$ and
$r$ copies of $\hat b$. Its length is therefore $k=(m+2)r$.
Commutation changes scalar factors but not these total degrees, so
no other power can have a nonzero trace. Expanding the logarithm in
\eqref{eq:period-trace-log} gives the second line of
\eqref{eq:period-moments}.

A closely related relation between quantum $A$-periods and traces of
lattice Hamiltonian powers is proposed in Section~3 of \cite{Gan2025}.
The equivalence of constant-term trace and wavefunction definitions of
the quantum $A$-period is established for the mirror-curve construction
of \cite{Grafnitz:2025anq}. The factorization~\eqref{eq:quantum-factorization}
gives a direct proof of the trace-log identity for the present family,
with the normalization needed below for the TBA relation.

At $q=1$, all words with the required numbers of the three monomials
have weight one. Their multinomial count gives the classical check
\[
 \left.\tau\!\left(\hat H^{(m+2)r}\right)\right|_{q=1}
 =\frac{((m+2)r)!}{(mr)!\,(r!)^2},
 \qquad
 \left.\Pi_m(z)\right|_{q=1}
 =\sum_{r\geq1}\frac{((m+2)r-1)!}{(mr)!\,(r!)^2}\,z^r.
\]
This verifies the normalization in~\eqref{eq:period-moments}.

\subsection{The change of quantum variables}

The change of variables is motivated by its classical limit. At $q=1$,
take commuting variables $a,b$ and put $U=a/b$, $Y=a+b$. Then
$a=UY/(1+U)$ and $b=Y/(1+U)$, so
\[
 a+b+a^{-m}b^{-1}
 =Y+U^{-m}(1+U)^{m+1}Y^{-(m+1)}.
\]
This leaves only the two powers $Y$ and $Y^{-(m+1)}$, which will
generate the two lattice moves. Its quantum counterpart is obtained
by defining
\begin{equation}\label{eq:quantum-change}
 \hat U=q^{(m+2)/2}\hat a\hat b^{-1},
 \qquad \hat Y=\hat a+\hat b.
\end{equation}
Direct calculation gives $\hat U\hat Y=q\hat Y\hat U$. We claim that
\begin{equation}\label{eq:mutated-curve}
 \hat H=\hat Y+P(\hat U)\hat Y^{-(m+1)},
\end{equation}
where $P$ is precisely the polynomial in \eqref{eq:P-definition}.
Here and below the order of the two factors in the second term matters.

The powers and shifts in \eqref{eq:mutated-curve} can be checked
explicitly. Repeated commutation gives
\begin{equation}\label{eq:U-negative-power}
 (\hat a\hat b^{-1})^{-m}
 =q^{m(m+1)/2}\hat a^{-m}\hat b^m,
 \qquad
 \hat U^{-m}=q^{-m/2}\hat a^{-m}\hat b^m.
\end{equation}
Moreover,
\[
 \hat Y=(1+q^{-(m+2)/2}\hat U)\hat b,
 \qquad \hat b^{-1}f(\hat U)=f(q\hat U)\hat b^{-1}.
\]
Moving the powers of $\hat b^{-1}$ through $\hat Y^{m+1}$ therefore gives
\begin{equation}\label{eq:quantum-product}
 \hat b^{-(m+1)}\hat Y^{m+1}
 =\prod_{j=1}^{m+1}(1+q^{j-(m+2)/2}\hat U).
\end{equation}
All factors in this product commute because they are functions of
$\hat U$ alone. Combining \eqref{eq:U-negative-power} and
\eqref{eq:quantum-product},
\begin{equation}
 \left(q^{-m/2}\hat a^{-m}\hat b^{-1}\right)\hat Y^{m+1}
 =\hat U^{-m}\prod_{j=1}^{m+1}(1+q^{j-(m+2)/2}\hat U)
 =P(\hat U).
\end{equation}
Multiplication on the right by $\hat Y^{-(m+1)}$ proves
\eqref{eq:mutated-curve}.

The constant-term trace is preserved by this change of variables.
To see this without assuming any invariance of quantum periods, first
use the pair $(\hat U,\hat b)$. A monomial
$\hat U^r\hat b^s$ is a nonzero scalar multiple of
$\hat a^r\hat b^{s-r}$, so its old constant term is nonzero only when
$r=s=0$, and is then one. Next, for every integer $s$,
\[
 \hat Y^s=g_s(\hat U)\hat b^s,\qquad g_0(\hat U)=1,
\]
where $g_s$ is a rational function regular at $\hat U=0$.
Negative powers are interpreted by expanding the inverses of
$1+q^j\hat U$ at $\hat U=0$; this also gives a precise algebraic meaning
to $\hat Y^{-1}$. If $s\ne0$, every term of
$\hat U^r g_s(\hat U)\hat b^s$ has nonzero $\hat b$-degree in the
$(\hat U,\hat b)$ variables and hence has zero old constant term.
If $s=0$, the constant term is one exactly when $r=0$. Thus
\begin{equation}\label{eq:trace-invariance}
 \tau(\hat U^r\hat Y^s)=\delta_{r,0}\delta_{s,0}.
\end{equation}
In particular, we may compute every trace in
\eqref{eq:period-moments} using the expression
\eqref{eq:mutated-curve} and extracting its $\hat U^0\hat Y^0$
coefficient.

\subsection{A finite lattice partition function}

The transformed operator has a simple lattice interpretation. On a
bi-infinite basis $\{e_i\}_{i\in\Z}$, represent it by
\begin{equation}\label{eq:lattice-representation}
 \hat U e_i=q^iX e_i,\qquad \hat Y e_i=e_{i+1}.
\end{equation}
Here $X$ is a commuting auxiliary variable. These operators satisfy
$\hat U\hat Y=q\hat Y\hat U$. The two terms in
\eqref{eq:mutated-curve} move a lattice point one step to the right or
$m+1$ steps to the left. The latter move, when it ends at $i$, has
weight $P(q^iX)$.

Restrict the matrix elements of this bi-infinite operator to the sites
$-M,\ldots,M$, where $M$ is a positive integer, omitting moves that
leave the interval. Denote the resulting $(2M+1)\times(2M+1)$ matrix
by $\hat H_M$. Explicitly,
\begin{equation}\label{eq:finite-matrix}
 (\hat H_M)_{ji}=\delta_{j,i+1}
       +P(q^jX)\delta_{j,i-m-1},\qquad -M\leq i,j\leq M.
\end{equation}
This is a restriction of the full difference operator; it does not
require an inverse of a finite shift matrix.

In the determinant expansion of $1-t\hat H_M$, every nontrivial
permutation cycle consists of $m+2$ consecutive sites. A permutation
cycle visits distinct sites until it returns to its starting point.
Choose its smallest site $i$ as the starting point. Before the first
leftward move, it follows rightward unit steps through
$i,i+1,\ldots,i+r$. The first leftward move lands at
$i+r-(m+1)$. If $r<m+1$, this is below the smallest site $i$.
If $r>m+1$, it revisits a site strictly between $i$ and $i+r$,
which is impossible in a permutation cycle before returning to $i$.
Thus $r=m+1$, and this move closes the cycle at $i$:
\[
 i\longrightarrow i+1\longrightarrow\cdots
 \longrightarrow i+m+1\longrightarrow i.
\]
The cycle therefore has exactly $m+1$ rightward steps and one
leftward step, and occupies precisely $m+2$ consecutive sites.

A cycle occupying $i,\ldots,i+m+1$ contributes
$-t^{m+2}P(q^iX)$: the permutation sign is $(-1)^{m+1}$ and the
$m+2$ off-diagonal entries supply $(-t)^{m+2}$.
Disjoint cycles are consequently equivalent to rods of length $m+2$
that cannot overlap. A rod starting at $i$ ends at $i+m+1$, so
its allowed starts satisfy $-M\leq i\leq M-m-1$.
With $z=t^{m+2}$, define their partition function by
\begin{equation}\label{eq:rod-partition}
 Z_M=\sum_{\substack{I\subset\{-M,\ldots,M-m-1\}\\
                |i-j|\geq m+2\ \text{for distinct }i,j\in I}}
          \prod_{i\in I}w_i,
 \qquad w_i=-zP(q^iX).
\end{equation}
The activity $w_i$ weights a rod starting at $i$. The empty
configuration gives $Z_M=1+O(z)$, so its inverse and logarithm are
well-defined formal series. The determinant argument proves
\begin{equation}\label{eq:rod-determinant}
 Z_M(X,t^{m+2})=\det(1-t\hat H_M).
\end{equation}
The activities need not be positive. This determinant interpretation
specializes the band-matrix construction for integer exclusion statistics in
\cite[Section~5.2]{OuvryPolychronakos2019}; see also equation~(30)
of the published version of~\cite{OuvryPolychronakos2021}.
Here the exclusion parameter is $m+2$, and the finite interval has
open boundaries.

The period is the constant term of minus the logarithm of this
partition function per lattice site:
\begin{equation}\label{eq:period-rod-limit}
 \Pi_m(z)=\lim_{M\to\infty}
       \frac{1}{2M+1}\operatorname{CT}_X[-\log Z_M(X,z)].
\end{equation}
To justify this coefficientwise limit, expand
\begin{equation}
 -\log\det(1-t\hat H_M)
 =\sum_{k\geq1}\frac{t^k}{k}\Tr(\hat H_M^k).
\end{equation}
A path of $k$ steps means a sequence $i_0,i_1,\ldots,i_k$ with
\[
 i_j-i_{j-1}\in\{1,-(m+1)\},\qquad 1\leq j\leq k.
\]
Each step applies one summand of $\hat H$. Matrix multiplication gives
\[
 (\hat H_M^k)_{ii}
 =\sum_{i_1,\ldots,i_{k-1}=-M}^{M}
   \prod_{j=1}^{k}(\hat H_M)_{i_j i_{j-1}},
 \qquad i_0=i_k=i.
\]
Each nonzero summand is the weight of a closed $k$-step path from
$i$ to itself that stays inside the interval. Its weight is the
product of $P(q^{i_j}X)$ over the leftward moves; all rightward
moves have weight one. Unlike a permutation cycle, such a path may
revisit sites before it closes.

For example, for $m=1$ the three closed paths of length three are
\[
 \begin{aligned}
 &i\to i+1\to i+2\to i,
 &&\text{weight }P(q^iX),\\
 &i\to i+1\to i-1\to i,
 &&\text{weight }P(q^{i-1}X),\\
 &i\to i-2\to i-1\to i,
 &&\text{weight }P(q^{i-2}X).
 \end{aligned}
\]
All three contribute on the infinite lattice. On the finite interval,
only those staying within $-M,\ldots,M$ are retained; for instance,
at $i=-M$ only the first is allowed when $M\geq1$.

For fixed $k$, every intermediate site satisfies
$|i_j-i_0|\leq j(m+1)\leq k(m+1)$. Hence all paths from a site with
\[
 -M+k(m+1)\leq i\leq M-k(m+1)
\]
stay in the interval, so $(\hat H_M^k)_{ii}=(\hat H^k)_{ii}$ there.
If the coefficient of $\hat Y^0$ in the ordered expansion of
$\hat H^k$ is $h_k(\hat U)$, only this term contributes to a diagonal
entry, and $(\hat H^k)_{ii}=h_k(q^iX)$. Constant-term extraction
removes the shift $q^i$, giving
\[
 \operatorname{CT}_X(\hat H_M^k)_{ii}
 =\operatorname{CT}_X h_k(q^iX)
 =\operatorname{CT}_X h_k(X)
 =\tau(\hat H^k)
\]
at every such bulk site. The last equality uses
\eqref{eq:trace-invariance}.

At most $2k(m+1)$ endpoint sites can differ. For any fixed sequence
of relative moves, translating its starting site from $0$ to $i$
replaces $X$ by $q^iX$ in its weight, and hence leaves its constant
term unchanged. The permitted paths near an endpoint depend only
on the distance from that endpoint. Once the two endpoint regions
are sufficiently far apart, their total constant-term correction is
therefore independent of $M$. Thus
\[
 \operatorname{CT}_X\Tr(\hat H_M^k)
 =(2M+1)\tau(\hat H^k)+O(1)
 \qquad(M\to\infty,\ k\text{ fixed}),
\]
where the correction comes from a fixed number of endpoint sites.
Dividing by $2M+1$ proves
\begin{equation}\label{eq:trace-lattice-limit}
 \lim_{M\to\infty}\frac{1}{2M+1}
     \operatorname{CT}_X\Tr(\hat H_M^k)=\tau(\hat H^k).
\end{equation}
Together with the determinant identity and its logarithmic expansion,
\eqref{eq:trace-lattice-limit} gives
\[
 \lim_{M\to\infty}\frac{1}{2M+1}
     \operatorname{CT}_X[-\log Z_M(X,t^{m+2})]
 =\sum_{k\geq1}\frac{t^k}{k}\tau(\hat H^k)
 =\Pi_m(t^{m+2}),
\]
where the last equality is~\eqref{eq:period-moments}.
This proves~\eqref{eq:period-rod-limit} with $z=t^{m+2}$; each
coefficient uses a fixed $k$, so no analytic thermodynamic limit
is required.

\subsection{The rod density equation and the residue identity}

Treat the activities $w_i$ temporarily as independent variables and set
\begin{equation}\label{eq:rod-density}
 \rho_i=w_i\frac{\partial\log Z_M}{\partial w_i}.
\end{equation}
Thus $\rho_i$ is the total weight of configurations containing a rod
starting at $i$, divided by $Z_M$. Activities and densities outside
$\{-M,\ldots,M-m-1\}$ are zero.

We first derive an exact identity on the finite interval. Let
$Z^L_j$ be the partition function with allowed starts at most $j$,
and $Z^R_j$ the one with allowed starts at least $j$, always intersected
with the original finite set of starts. Fixing a rod at $i$ separates
the remaining starts into two independent intervals, giving
\begin{equation}\label{eq:density-factorization}
 \rho_i=w_i\frac{Z^L_{i-m-2}Z^R_{i+m+2}}{Z_M}.
\end{equation}
Define the two empty-window weights
\begin{equation}\label{eq:empty-window-definitions}
 C_j=1-\sum_{k=j}^{j+m+1}\rho_k,
 \qquad D_j=1-\sum_{k=j}^{j+m}\rho_k.
\end{equation}
Each window contains at most one occupied rod start in any
configuration. Subtracting the weights for these mutually exclusive
starts leaves the normalized weight for no start in the window.
The remaining intervals are independent, so
\begin{equation}\label{eq:empty-window-factorizations}
 C_j=\frac{Z^L_{j-1}Z^R_{j+m+2}}{Z_M},
 \qquad
 D_j=\frac{Z^L_{j-1}Z^R_{j+m+1}}{Z_M}.
\end{equation}
Taking products makes all intermediate factors cancel:
\begin{equation}\label{eq:window-product-ratio}
 \frac{\displaystyle\prod_{j=i-m-1}^{i}C_j}
      {\displaystyle\prod_{j=i-m}^{i}D_j}
 =\frac{Z^L_{i-m-2}Z^R_{i+m+2}}{Z_M}.
\end{equation}
These divisions are legitimate in formal series because every
partition function and empty-window weight has constant term one.
Using \eqref{eq:density-factorization}, we obtain the exact density
equation
\begin{equation}\label{eq:rod-density-equation}
 \rho_i\prod_{j=i-m}^{i}D_j
 =w_i\prod_{j=i-m-1}^{i}C_j.
\end{equation}
This is equation~(36) in the published version of
\cite{OuvryPolychronakos2021}, specialized to exclusion parameter
$m+2$, occupation $\rho_i$, and activity $w_i$.
The empty-window factorizations and cancellation above give its
Appendix~A derivation in the present notation.

The required locality bound follows directly from
\eqref{eq:rod-density}. Keeping the activities independent, consider
a monomial $w_{j_1}^{n_1}\cdots w_{j_s}^{n_s}$ in $\log Z_M$, where
$j_1<\cdots<j_s$, $n_a\geq1$, and $\sum_a n_a=r$.
Setting all other activities to zero leaves its coefficient unchanged.
If some gap $j_{a+1}-j_a\geq m+2$, the remaining rods separate into
two independent groups. The restricted partition function factorizes
as $Z_{\mathrm{left}}Z_{\mathrm{right}}$, so its logarithm is a sum
and contains no monomial involving both groups. Thus a nonzero
coefficient requires every successive gap to be at most $m+1$.
The operation $w_i\partial_{w_i}$ retains only monomials containing
$i$ and preserves their degree. Every start $j_a$ in such a term
therefore satisfies
\[
 |j_a-i|\leq j_s-j_1\leq(s-1)(m+1)\leq(r-1)(m+1).
\]
After specializing $w_j=-zP(q^jX)$, degree $r$ becomes order $z^r$.
Hence $[z^r]\rho_i$ depends only on activities within this distance
of $i$.

Enlarging the allowed-start interval beyond this neighborhood leaves
these coefficients unchanged, defining a bulk formal series at every
fixed $i$. At site $0$, the coefficient of $z^r$ stabilizes for
$M\geq r(m+1)$. The finite density equation
\eqref{eq:rod-density-equation} therefore also holds in the bulk.

The site dependence follows from the identity
\[
 w_{i+d}(X;q,z)=-zP(q^{i+d}X;q)=w_d(q^iX;q,z).
\]
Since the exclusion rule depends only on separations, the same change
of $X$ relates the local problems at sites $i$ and $0$. Thus, in the
bulk,
\[
 \rho_i(X;q,z)=\rho(q^iX;q,z),\qquad \rho(X;q,z):=\rho_0(X;q,z).
\]
We again suppress the fixed parameters $q,z$; hereafter $\rho_i$
denotes the bulk density unless a finite interval is specified.
Define
\begin{equation}\label{eq:eta-density}
 \eta_m(q^iX)=1-(m+2)\rho_i.
\end{equation}
Equations \eqref{eq:second-AB} and
\eqref{eq:empty-window-definitions} then give
\begin{equation}\label{eq:AB-windows}
 A(q^jX)=(m+2)C_j,\qquad B(q^jX)=(m+2)D_j.
\end{equation}
Substituting \eqref{eq:eta-density} and \eqref{eq:AB-windows} into
\eqref{eq:rod-density-equation}, all powers of $m+2$ cancel and we find
\begin{equation}
 (1-\eta_m(q^iX))\prod_{j=i-m}^{i}B(q^jX)
 =-zP(q^iX)\prod_{j=i-m-1}^{i}A(q^jX).
\end{equation}
Setting $i=0$ in this bulk equation gives exactly
\eqref{eq:second-TBA}. Its unique solution
with leading term one is therefore the function constructed from the
rod density.

Finally, all activities are proportional to $z$, so on every finite
interval
\begin{equation}\label{eq:density-log-derivative}
 \theta_z[-\log Z_M(X,z)]=-
     \sum_{i=-M}^{M-m-1}\rho_i.
\end{equation}
At any fixed order in $z$, only finitely many endpoint densities differ
from the bulk density. Constant-term extraction removes the shift
$q^i$ from each bulk contribution. The number of allowed rod starts
is $2M-m$, and $(2M-m)/(2M+1)\to1$. Dividing
\eqref{eq:density-log-derivative} by $2M+1$, taking the constant term, and
then taking the coefficientwise limit gives
\begin{equation}\label{eq:period-density}
 \theta_z\Pi_m(z)=-\operatorname{CT}_X\rho(X).
\end{equation}
The logarithmic derivative commutes with this limit because it simply
multiplies the coefficient of $z^r$ by $r$. Taking the constant term
of \eqref{eq:eta-density} now proves
\begin{equation}
 \operatorname{CT}_X\eta_m(X)
 =1-(m+2)\operatorname{CT}_X\rho(X)
 =1+(m+2)\theta_z\Pi_m(z),
\end{equation}
as claimed. The factor $m+2$ is fixed by the rod length and the
normalization in~\eqref{eq:eta-density}.

\section{Conclusion}\label{sec:conclusion}

We have proved the quadratic and higher-degree quantum period--TBA
relations as formal series in $z$, retaining the full dependence on
$q=e^{i\hbar}$. The quadratic proof compares constant terms of
oppositely shifted products. The higher-degree proof identifies the
quantum period with a cyclic quantum-torus trace and then with the
free energy of nonoverlapping rods. Their density equation explains
the TBA-like equation and its normalization, accounting for the same
period in two different TBA descriptions.

Completing these identities into a physical spectral description
requires analytic continuation in $z$, Stokes data, and boundary
conditions. The constant-term trace and finite determinants used here
are not automatically the operator trace and Fredholm determinant of
the physical inverse mirror-curve operator. Establishing their relation
would clarify the nonperturbative effects discussed in
\cite{Kallen:2013qla,Grassi:2014zfa,Wang:2015wdy}.
A related task is to obtain quantum $B$-periods and global connection
data for exact quantization. The exact-WKB and wall-crossing approaches
of \cite{Ito:2018eon,DelMonteLonghi2023,Ito:2024wxw,Shu:2025mwe}
provide possible starting points.

For more general mirror curves and several dynamical moduli,
additional monomials introduce several lattice moves, and the
exclusion model may require rods of different lengths. Deriving its
density equations could help describe the independent periods entering
higher-genus quantization conditions
\cite{Franco:2015rnr,Huang:2020neq}. The Weyl symmetries of del Pezzo
quantum curves \cite{Moriyama:2020lyk} may constrain this extension
and the action of quantum-variable changes on the periods.

The enumerative interpretation of quantum periods
\cite{Grafnitz:2025anq} also suggests seeking a geometric meaning for
the rods and their $q$-dependent weights. For local $\mathbb P^2$,
one can compare this model, after the appropriate mirror-map
transformation, with logarithmic Gromov--Witten invariants. Beyond the
geometric hypotheses of that work, the appropriate curve-counting
theory and boundary contributions remain to be identified. Such a
comparison may also clarify the integrality and BPS-state interpretation
of the resulting series.

\vspace{0.2in} {\leftline {\bf Acknowledgments}}
\nopagebreak
The derivations and proofs in this paper were developed with assistance from GPT-6 Astra. The text was written by the author with assistance from AI tools. The author is responsible for checking the arguments and for the final manuscript. We thank Di Yang for helpful discussions. This work is  supported by the National Natural Science Foundation of China Grants No. 12325502 and No. 12247103.  

\appendix

\phantomsection
\addcontentsline{toc}{section}{References}

\IfFileExists{utphys.bst}{\bibliographystyle{utphys}}{\bibliographystyle{unsrt}} 
\bibliography{ReferenceTBA}


\end{document}